\documentclass[reprint,prl,aps,a4paper,10pt,titlepage]{revtex4-1}
\usepackage{ucs}
\usepackage[utf8x]{inputenc}
\usepackage{amsmath}
\usepackage{amsfonts}
\usepackage{amssymb}
\usepackage{amsthm}
\usepackage[english]{babel}
\usepackage[T1]{fontenc}
\usepackage{graphicx}
\usepackage[dvips]{hyperref}

\begin{document}
\title{Constraints on the Dark Photon Parameter Space from Leptonic Rare Kaon Decays
}
\author{T.~Beranek}
\author{M.~Vanderhaeghen}
\affiliation{Institut f\"ur Kernphysik, Johannes Gutenberg-Universit\"at, D-55099 Mainz, Germany}
\date{\today}

\newcommand{\fslash}[1]{\gamma\cdot{#1}}
\newcommand{\Abs}[1]{|\vec{#1}\,|}
\newcommand{\Absp}[1]{|\vec{#1}^{\,\prime}|}
\newcommand{\vecp}[1]{\vec{#1}^{\,\prime}}
\newcommand{\ubar}{{\overline{u}}}
\newcommand{\vbar}{{\overline{v}}}
\newcommand{\Ap}{A^\prime}

\begin{abstract}
With anomalies found in cosmic ray observations and unsolved questions of the Standard Model of particle physics like the discrepancy in the muon's anomalous magnetic moment, the idea of an $U(1)$ extension of the Standard Model arose. This extension allows for the interaction of Dark Matter by exchange of a photon-like massive force carrier $\Ap$ not included in the Standard Model. We discuss the possibility to constrain the dark photon parameter space by using data taken from rare kaon decays. Therefore we analyze the decay $K\rightarrow \mu \nu \Ap$ as a signal process and calculate the expected Standard Model background. Using these results we calculate new limits on the $\Ap$ parameter space, providing  motivations for new rare kaon decay experiments to extend the existing $\Ap$ bounds.
\end{abstract}

\maketitle

\begin{figure}
 \includegraphics[width=.48\linewidth]{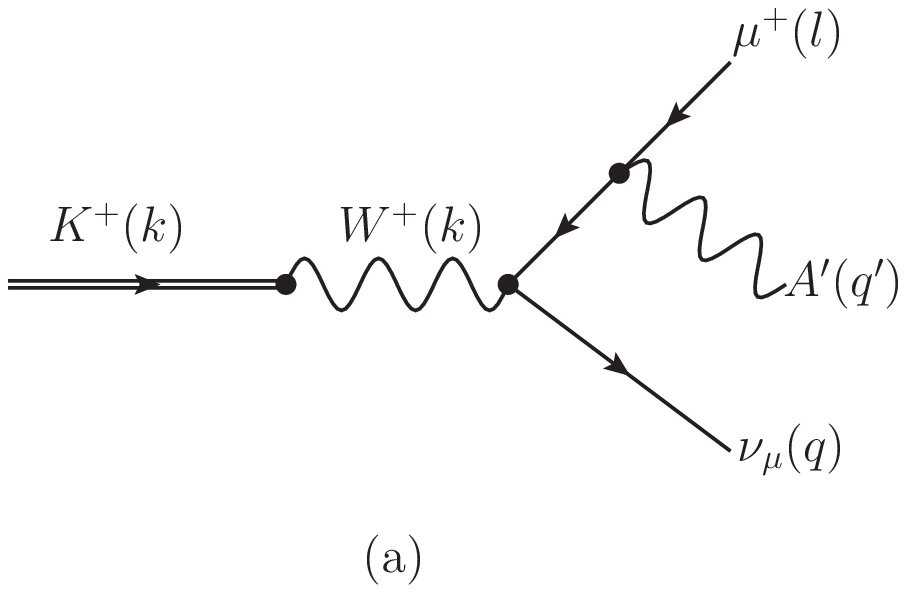}
\includegraphics[width=.48\linewidth]{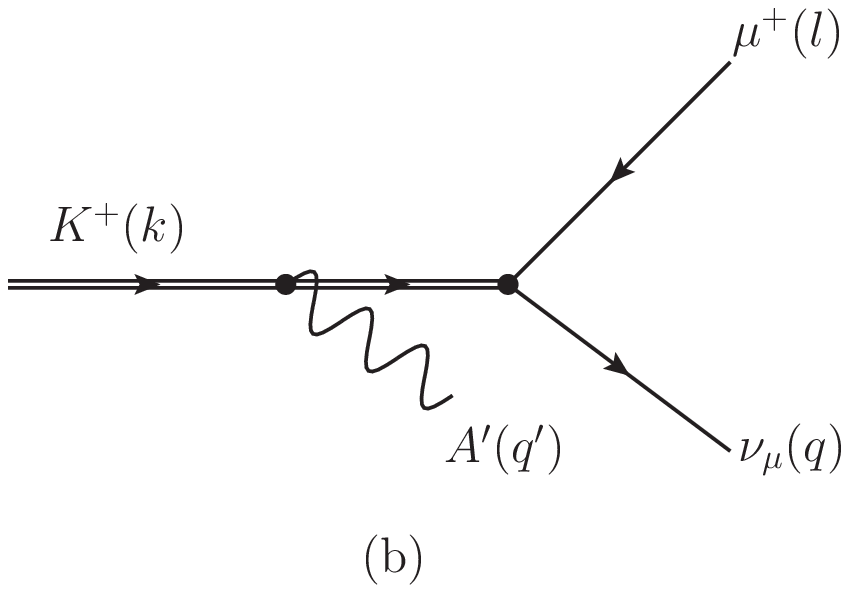}\\
\centering\includegraphics[width=.48\linewidth]{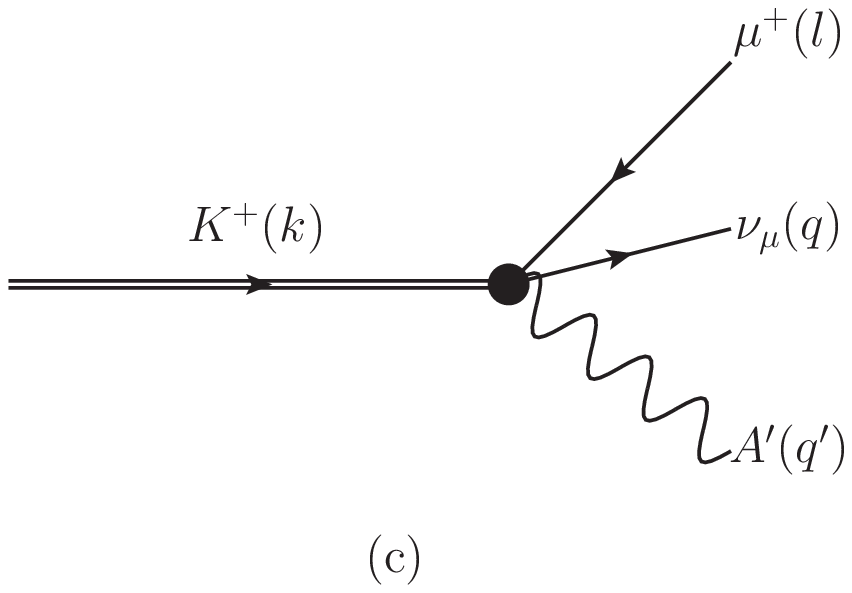}
\caption{\label{fig:feyn_Ap_sig}Feynman diagram for the process $K^+\rightarrow \mu^+ \nu_\mu \Ap$.}
\end{figure}

\begin{figure}[t]
\includegraphics[width=.48\linewidth]{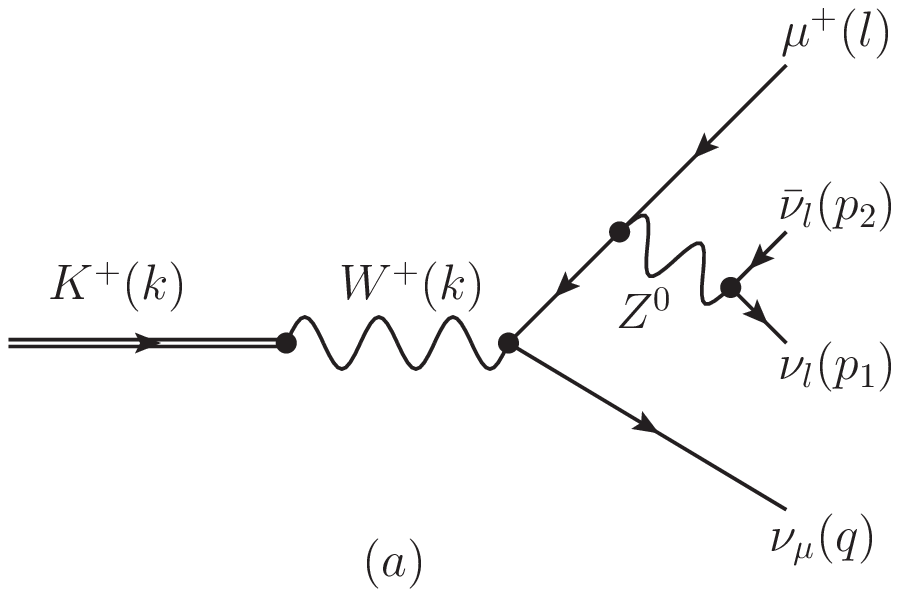}
\includegraphics[width=.48\linewidth]{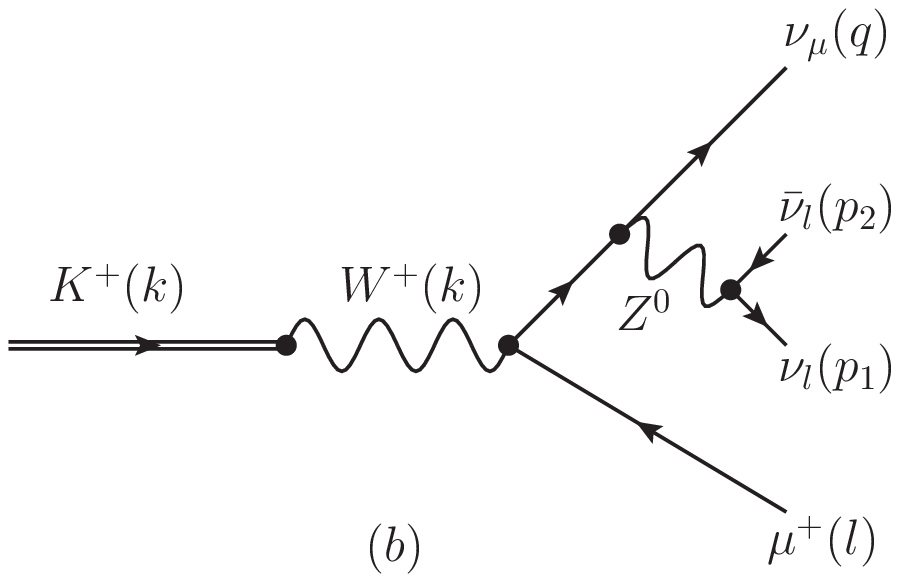}\\
\includegraphics[width=.48\linewidth]{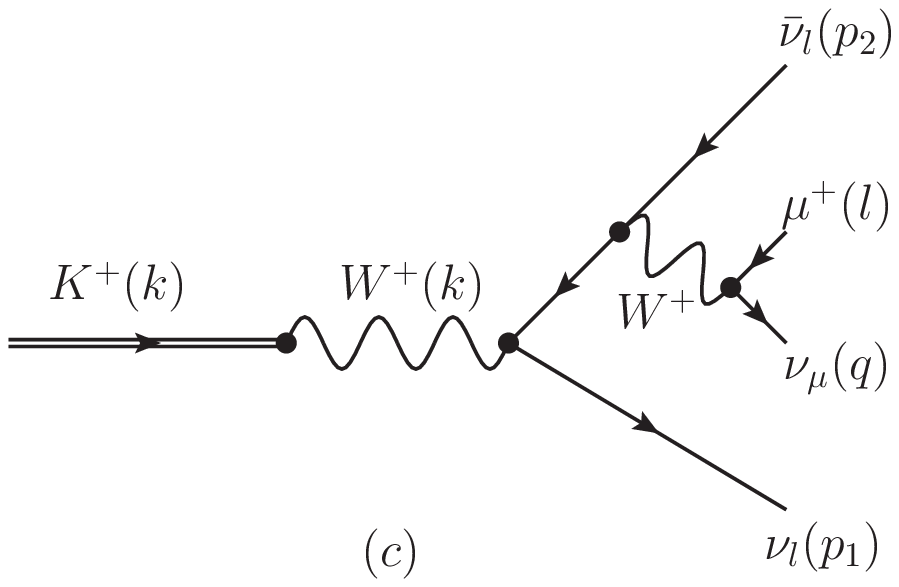}
\includegraphics[width=.48\linewidth]{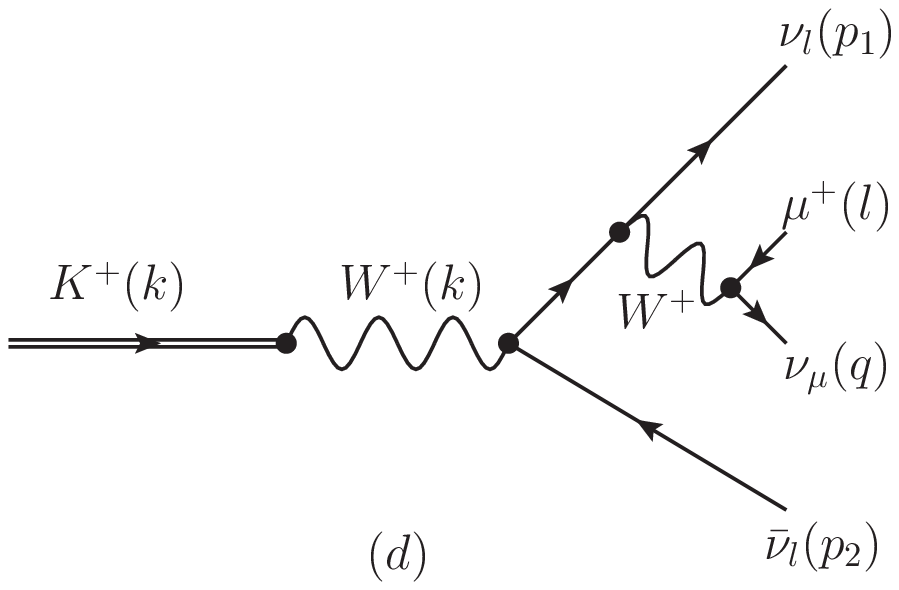}
\caption{\label{fig:feyn_mu3nu_bkg}SM background from $K^+\rightarrow \mu^+ \nu_\mu \nu_l\bar{\nu}_l$.}
\end{figure}

The possibility to explain anomalies in astrophysics and particle physics \cite{Strong:2005zx,Adriani:2008zr,Cholis:2008wq} by extending the Standard Model of Particle Physics (SM) by an additional $U(1)_D$ gauge group manifesting itself in a massive gauge boson $\Ap$ (``Dark Photon'') in the MeV to GeV mass range \cite{Fayet:1990wx,Holdom:1986eq,ArkaniHamed:2008qn} motivated a strong activity in theoretical as well as in experimental physics \cite{Pospelov:2008zw,Bjorken:2009mm,Essig:2009nc,Merkel:2011ze,Abrahamyan:2011gv,Archilli:2011zc,Blumlein:2011mv,Batell:2011qq,Carlson:2012pc,Aditya:2012ay}. In a recent work \cite{Carlson:2012pc}, where new physics scenarios were explored as a possible explanation of the proton charge radius problem, the constraints from rare kaon decays were examined. In the present work, we will use these rare kaon decays to find constraints on the $\Ap$ parameters. The $\Ap$ couples to the electromagnetic current via kinetic mixing giving rise to a QED-like vertex term
\[
 i \varepsilon \,e \, \gamma_\mu,
\]
where $\varepsilon$ is the kinetic mixing parameter describing the coupling strength to the electromagnetic current by $\varepsilon^2~=~\alpha^\prime_D/\alpha_\text{QED}$ with $\alpha_\text{QED}\equiv e^2/4\pi$ \cite{Bjorken:2009mm}.
\\
In the following we will study the process $K^+\rightarrow \mu^+\nu_\mu \Ap$ as a possible signal from the dark sector (see Feynman diagram in Fig.~\ref{fig:feyn_Ap_sig}) within the mentioned framework of kinetic mixing (model I) as well as in a model where the $\Ap$ couples only to the muon assuming an explicit breaking of gauge invariance applied in Ref.~\cite{Batell:2011qq} (model II). In the pioneering experiment \cite{Pang:1989ut} of the decay $K^+\rightarrow \mu^+ + \text{ neutrals,}$ only the charged muon is detected, excluding further charged particles or photons in the final state. Therefore this process cannot be distinguished from the $K^+\rightarrow \mu^+ \nu_\mu \nu_l\bar{\nu}_l$ decay (Feynman diagrams in Fig.~\ref{fig:feyn_mu3nu_bkg}), making it crucial to obtain a precise understanding of the background. This analysis will allow us to compute bounds on the $\Ap$ coupling strength $\varepsilon$ as a function of its mass parameter $m_{\Ap}$.\\

In calculating the decays of Figs.~\ref{fig:feyn_Ap_sig},~\ref{fig:feyn_mu3nu_bkg} we will use, that the largest external momentum scale of the considered processes, the kaon mass $m_K$, is far below the weak gauge boson masses $m_W$ and $m_Z$ allowing to approximate the weak interaction processes by point processes, respectively. Furthermore we parametrize the hadronic weak $K^+$ decay current by $\big<0\big|\bar{v}_s\gamma_\alpha(1-\gamma_5)u_u\big|K^+(k)\big>=f_K k_\alpha$, where $f_K$ is the kaon decay constant.\\
Since the $K^+$ is assumed to decay in rest the analysis will be performed in the $K^+$ rest frame where the kaon four-vector is given as $k=(m_K,\,\vec{0})$. Furthermore we use the momentum four-vectors $l = (E_\mu,0,0,\Abs{l})$ for the muon, $q = (E_\nu,E_\nu\,\sin \theta_\nu,0,E_\nu\,\cos \theta_\nu)$ for the neutrino, and $q' = k - l - q$ for the $\Ap$, where we choose the z-axis along the muon 3-momentum. All other 3-momenta and all angles associated with other particle are not observable and therefore these decays have isotropic muon angular distributions.\\
The $\Ap$ production amplitude derived from diagram~\ref{fig:feyn_Ap_sig} (a) is given by
\begin{align}
 \mathcal{M}_{\Ap,a} &= \frac{G_F\,f_K\,\varepsilon e\,\sin \theta_c}{\sqrt{2}\left((k-q)^2-m_\mu^2 \right)}\,\varepsilon^\ast_\alpha (q')  \label{eq:amp_Ap}\\
  &\quad\times\left[ \bar{u}(q)(1+\gamma_5)\left((k-q)^2+m_\mu \fslash{k} \right) \gamma^\alpha v(l) \right],\nonumber
\end{align}
where $G_F$ is the Fermi constant and $\theta_c$ the Cabbibo mixing angle. For model II the signal process is completely described by this amplitude. Due to the need for gauge invariance within the kinetic mixing framework, in model~I the amplitude is given by the coherent sum over all diagrams of Fig.~\ref{fig:feyn_Ap_sig} 
\begin{align}
 \mathcal{M_{\Ap}} &= \frac{G_F\,\varepsilon e\,\sin \theta_c}{\sqrt{2}}\,\varepsilon^\ast_\rho (q')\,\left(f_K\, m_\mu \,L^\rho - H^{\rho\nu} j_\nu \right),\label{eq:amp_Ap2}
\end{align}
\begin{align}
L^\rho &=\bar{u}(q)(1+\gamma_5)\left\{ \frac{2k^\rho-q^{\prime \rho}}{2k\cdot q^\prime - q^{\prime 2}} - \frac{2l^\rho-\fslash{q^\prime}\gamma^\rho}{2l\cdot q^\prime + q^{\prime 2}}\right\} v(l), \nonumber\\
j_\nu &= \bar{u}(q)\gamma_\nu (1+\gamma_5) v(l),\nonumber\\
H^{\rho \nu} &= -i\,V_1 \varepsilon^{\rho \nu \alpha\beta} q^\prime_\alpha k_\beta - A_1 (q^\prime \cdot W - W^\rho q^{\prime \nu})\nonumber\\
&\quad - A_2 (q^{\prime 2}g^{\rho\nu}-q^{\prime \rho}q^{\prime \nu})\nonumber,
\end{align}
whereas $W=k-q^\prime$ and $\varepsilon^{0123}=1$. The term proportional to $f_K$ is known as inner Bremsstrahlung contribution (IB) and does not contain any structure effects. The contribution proportional to $H^{\rho \nu}$ contains the structure dependent terms, which are parametrized by the form factors $V_1$, $A_1$, and $A_2$.\\
In the $K^+$ rest frame the differential decay width for $K^+\rightarrow \mu^+\nu_\mu \Ap$ using the conventions of Ref.~\cite{Bijnens:1992en} then reads as
\begin{align}
 \frac{d\Gamma(K^+\rightarrow \mu^+\nu_\mu \Ap)}{dE_\mu \, dE_\nu} &= \frac{1}{64\pi^3 m_K}\overline{\left|\mathcal{M_{\Ap}} \right|^2},\label{eq:dgamma_Ap}
\end{align}
where we have fixed the angle $\theta_\nu$ by evaluating the energy conserving $\delta$-function as:
\[
 \cos \theta_\nu = \frac{m_K^2+m_\mu^2-m_{\Ap}^2+2E_\mu E_\nu - 2m_K(E_\mu + E_\nu)}{2E_\nu \Abs{l}}.
\]
To obtain the decay width, Eq.~(\ref{eq:dgamma_Ap}) has to be integrated over $E_\mu$ and $E_\nu$ within the limits as given in section IV of Ref.~\cite{Carlson:2012pc}.
The background contribution in terms of its 5-fold differential decay rate is given by
\begin{align}
\begin{split}
 &\frac{d \Gamma (K^+\rightarrow \mu^+ \nu_\mu \nu_l\bar{\nu}_l)}{dE_\mu\, dE_\nu\,d\cos \theta\, d\Omega_1}\\
&\quad= \frac{1}{16m_K}\,\frac{1}{(2\pi)^6}\,\frac{\Abs{l}\Abs{q}\Abs{p_1}}{\left|E_1+E_2+\left(\vec{l}+\vec{q} \right)\cdot \hat{p}_1 \right|} \,\overline{\left|\mathcal{M} \right|^2},\label{eq:dgamma_mu3nu}
\end{split}
\end{align}
in the kaon rest frame, where
\[
 E_1 = \frac{m_K^2+m_\mu^2+2\,l\cdot q -2 m_K(E_\mu + E_\nu)}{2\left(m_K+(\vec{l}+\vec{q})\cdot \hat{p}_1 - E_\mu - E_\nu \right)}
\]
is fixed by the energy conserving $\delta$-function, $E_2=p_2^0$ and $\mathcal{M}$ is the coherent sum over the amplitudes derived from the Feynman diagrams in Fig.~\ref{fig:feyn_mu3nu_bkg}.\\
In order to obtain a dimensionless quantity, it is helpful to consider the ratio of these decay rates relative to the one for the $K^+\rightarrow \mu^+ \nu_\mu$ expressed by
\[
 \Gamma(K^+\rightarrow \mu^+ \nu_\mu) = \frac{G_F^2\,f_K^2\,\sin^2\theta_c}{8\pi\,m_K^3}\,m_\mu^2\,(m_K^2-m_\mu^2)^2.
\]
\begin{figure*}[!t]
\includegraphics[scale=.4,angle=270]{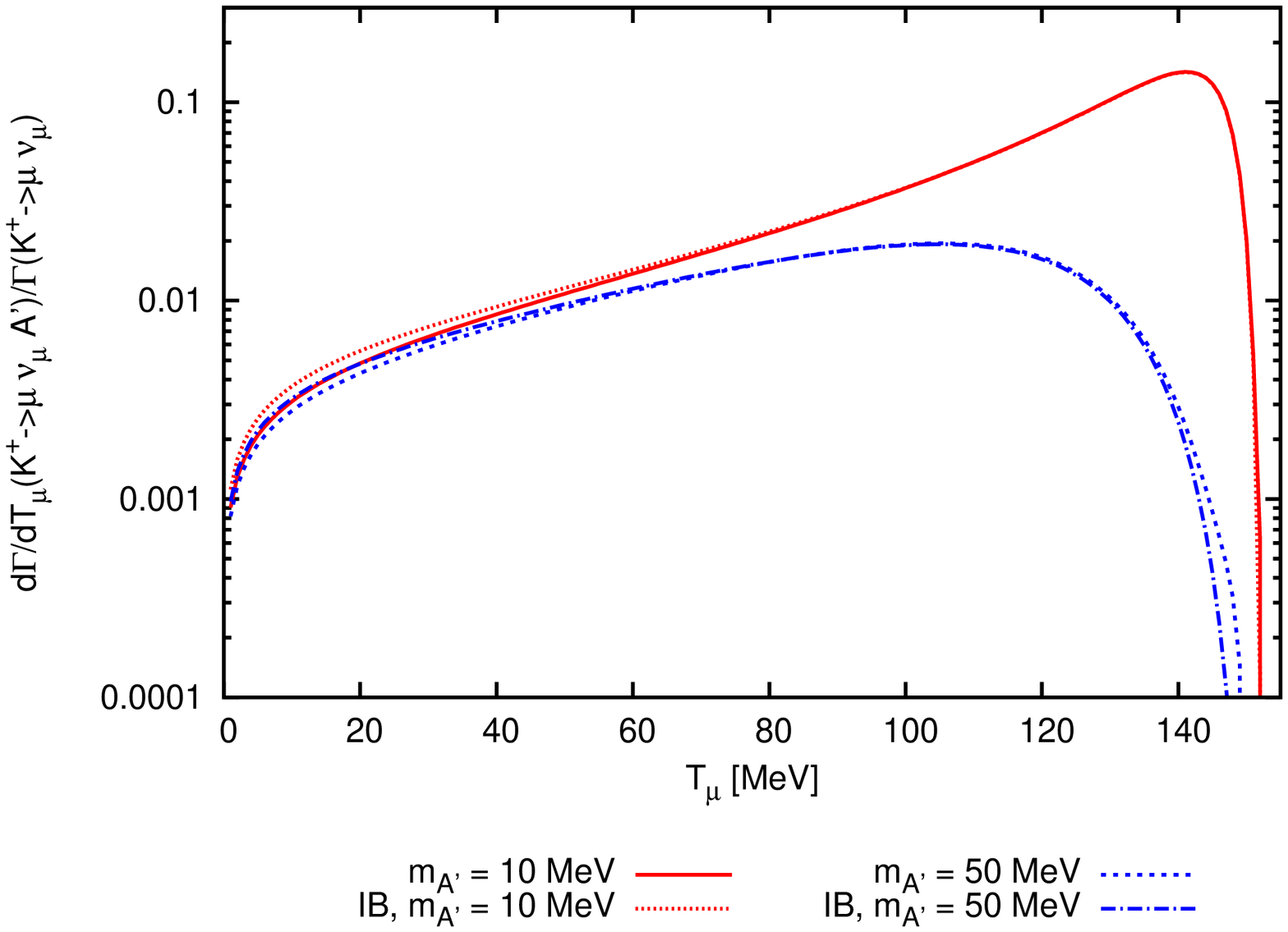}
\includegraphics[scale=.4,angle=270]{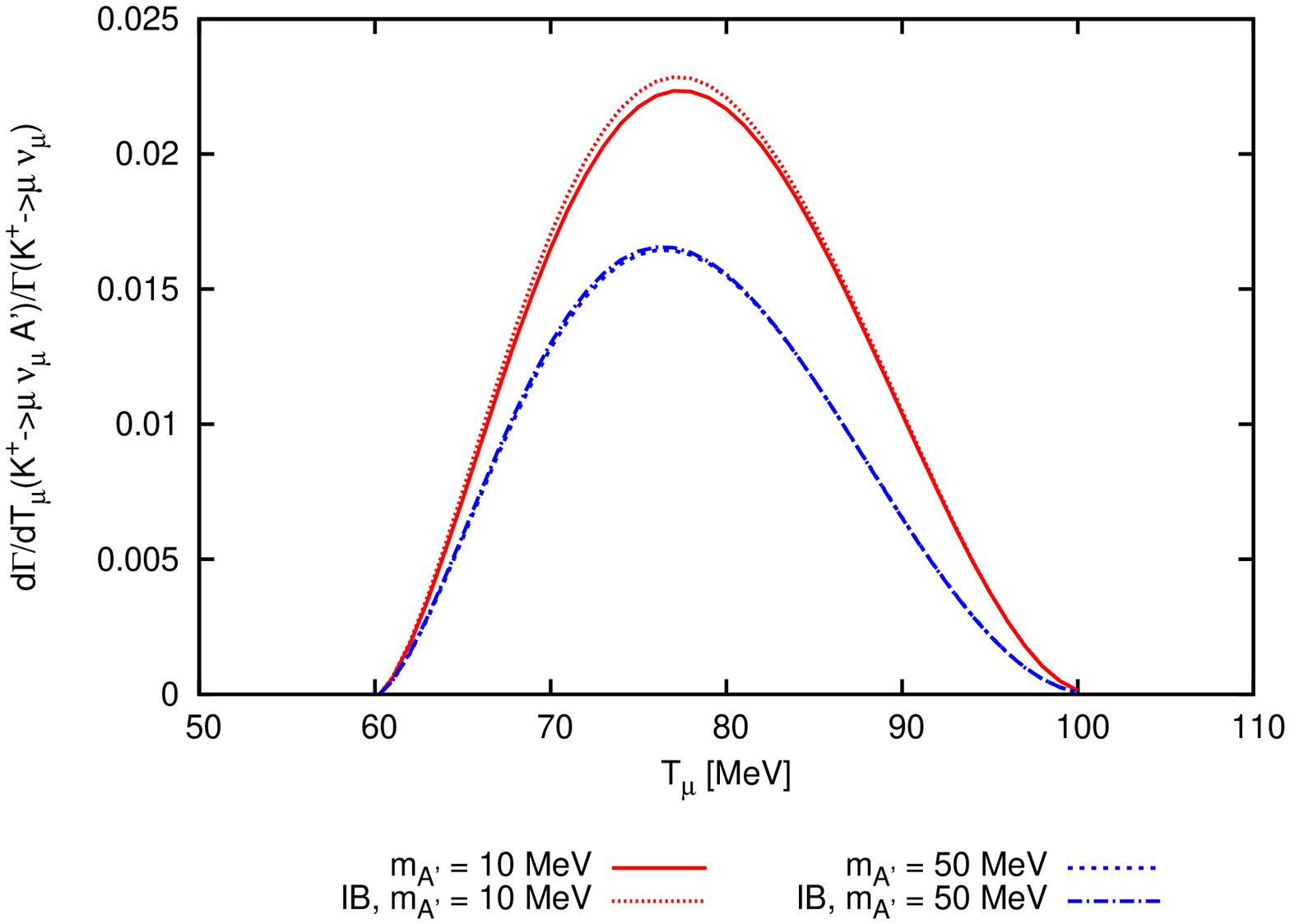}\\
\includegraphics[scale=.45,angle=270]{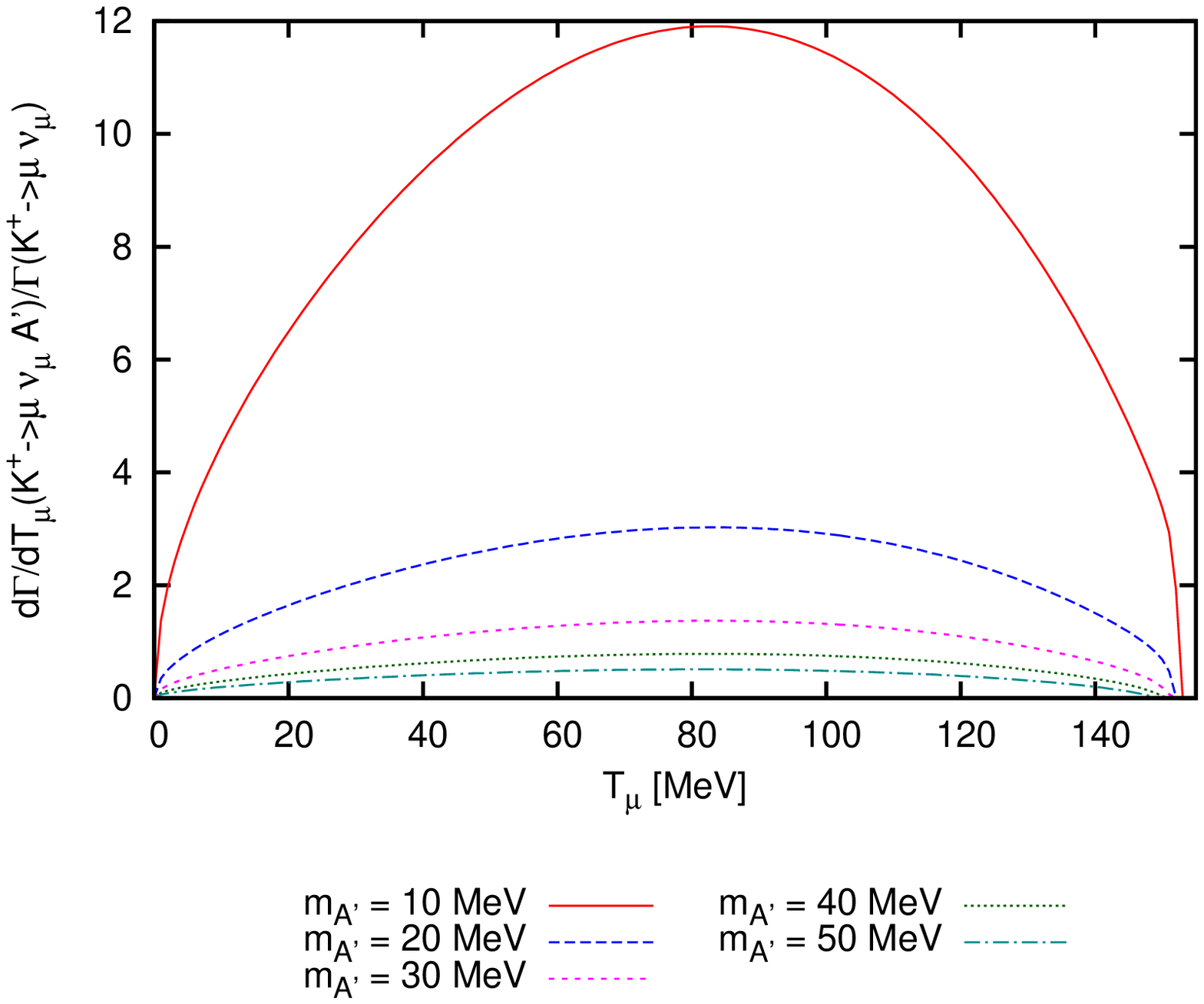}
\includegraphics[scale=.45,angle=270]{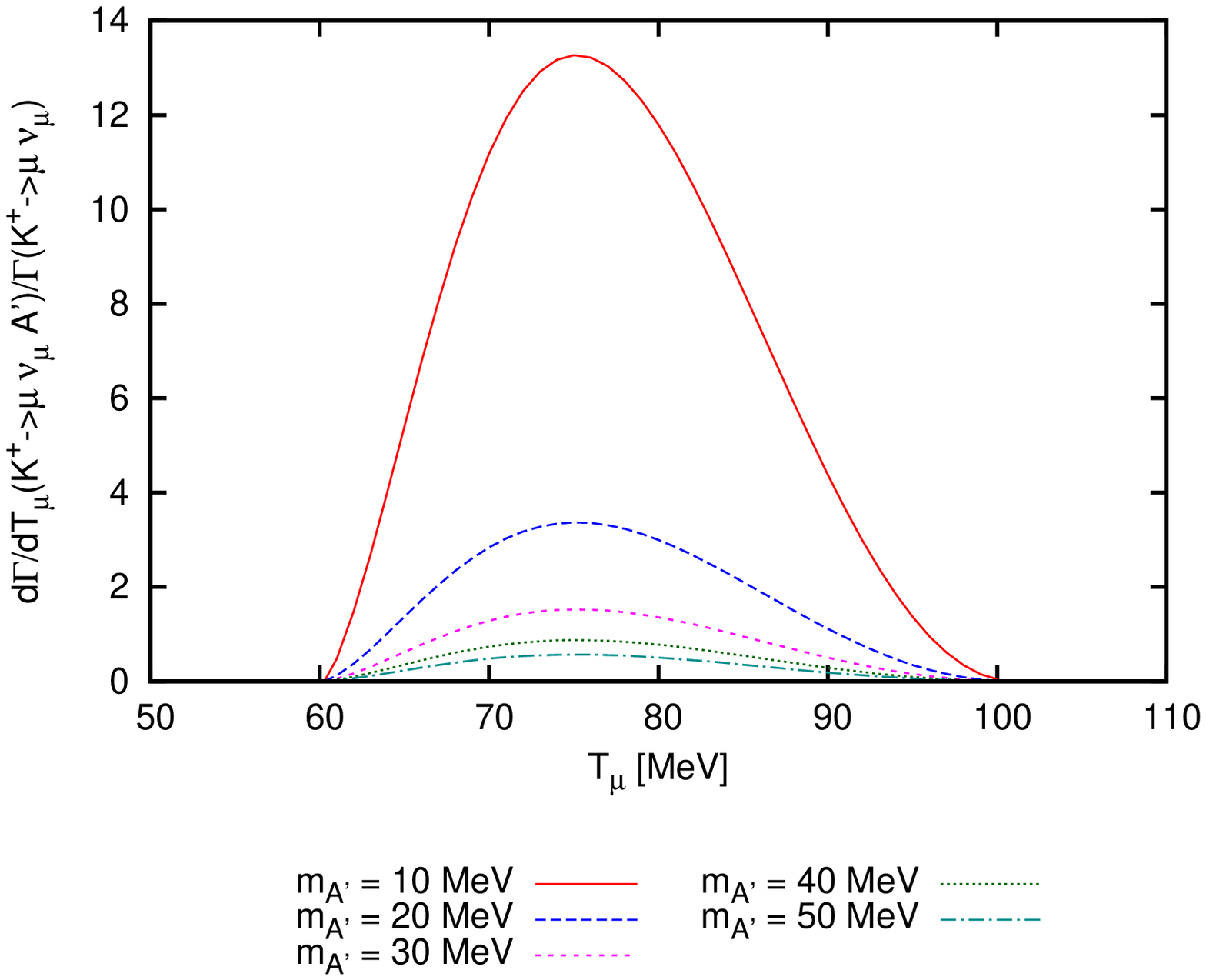}\\
\includegraphics[scale=.4,angle=270]{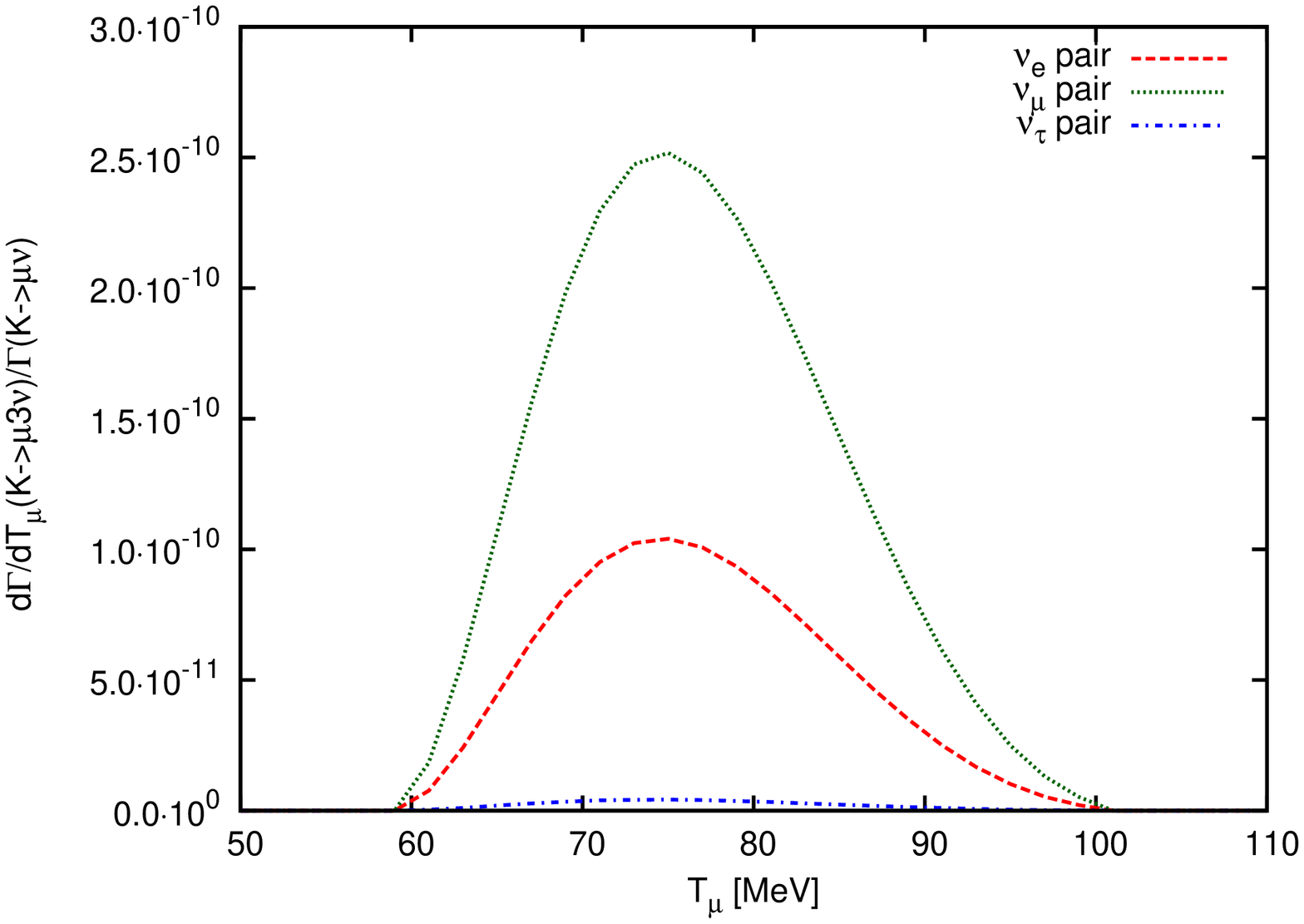}
\includegraphics[scale=.4,angle=270]{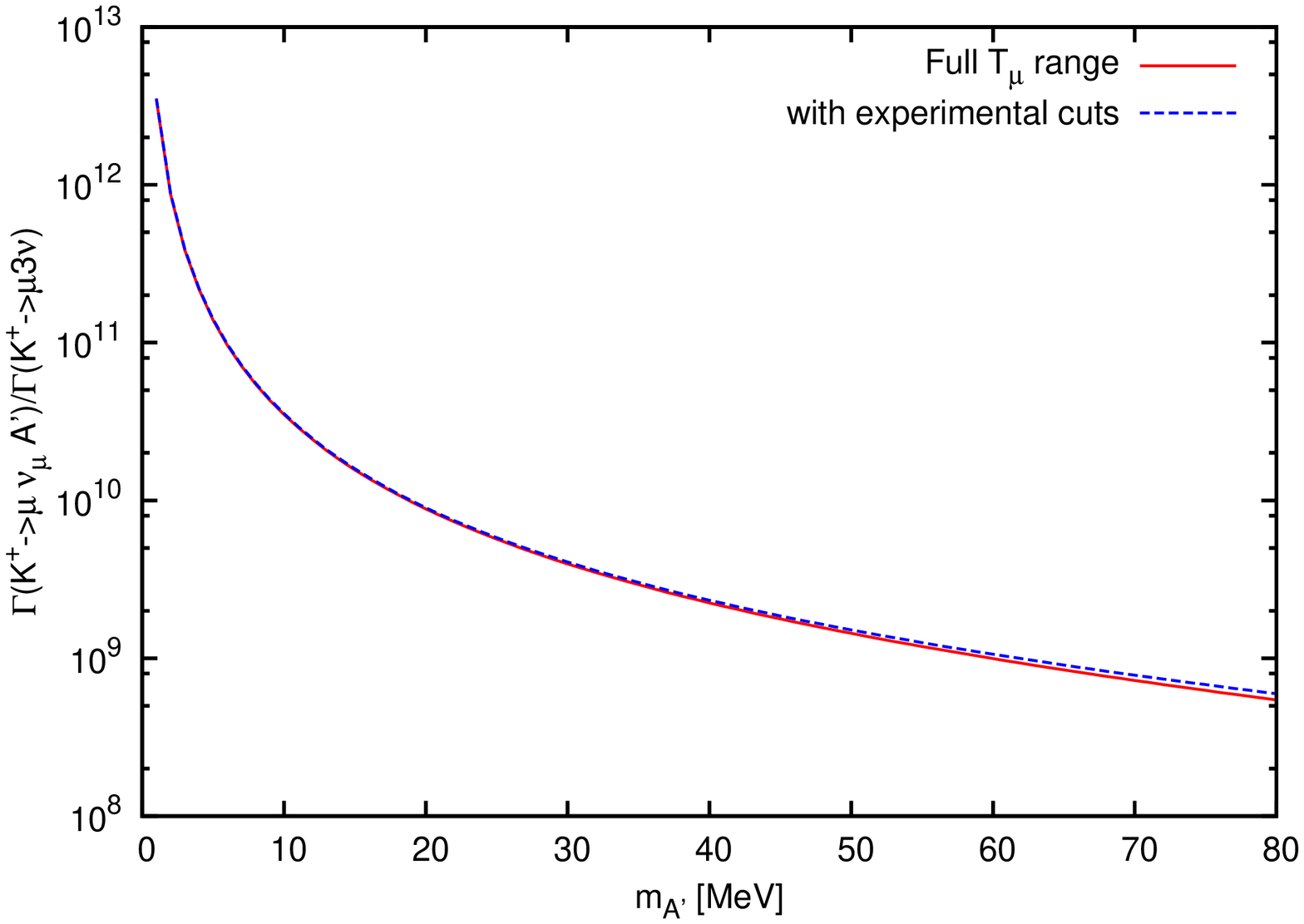}
\caption{\label{fig:decay_rate_plots}Upper and middle panels: Ratio of $\frac{d\Gamma}{dT_\mu}(K^+\rightarrow\mu^+\nu_\mu \Ap)$ and $\Gamma (K^+\rightarrow \mu^+ \nu_\mu)$ for various $\Ap$ masses for perfect detector efficiency (left panels) and for finite detector efficiency of Ref.~\cite{Pang:1989ut} (right panels) at $\varepsilon^2=1$. Upper panels: kinetic mixing model (model I); middle panels: model II, where the $\Ap$ only couples to the $\mu^+$. Lower left panel: Standard Model background for different neutrino families using the detector efficiency function. Lower right panel: ratio of total decay rates $\Gamma (K^+\rightarrow\mu^+\nu_\mu \Ap)$ at $\varepsilon^2=1$ relative to $\Gamma (K^+\rightarrow\mu^+\nu_\mu \nu\bar{\nu})$ in model II.}
\end{figure*}
Similar to the distributions (\ref{eq:dgamma_Ap}) and (\ref{eq:dgamma_mu3nu}) this decay rate is isotropic in the muon angles. Thus all corrections due to an angular detector acceptance being not $4\pi$ will cancel each other in the ratio with respect to the decay rate $\Gamma(K^+\rightarrow \mu^+ \nu_\mu)$.\\
For our analysis, we use the existing data published by Pang et al. in Ref.~\cite{Pang:1989ut}, who found an upper limit of $\Gamma(K^+\rightarrow \mu^+X)/\Gamma(K^+\rightarrow \mu^+ \nu_\mu) < 2 \cdot 10^{-6}$ on the ratio of the decay rates, where $X$ is a possible, not detectable neutral state not containing a photon.
In order to obtain the experimental limits from these data the differential decay rate $\frac{d \Gamma}{dE_\mu}(K^+\rightarrow \mu^+\nu_\mu \Ap)$ has to be folded with the detector efficiency $D(E_\mu)$ given in Ref.~\cite{Pang:1989ut}, i.e.
\begin{equation}
 \tilde{R} (m_{\Ap}):=\frac{\int \frac{d \Gamma}{dE_\mu}(K^+\rightarrow \mu^+\nu_\mu \Ap) D(E_\mu) dE_\mu}{ \Gamma(K^+\rightarrow \mu^+ \nu_\mu)} .
\end{equation}
Since the kinetic mixing factor $\varepsilon$ is a global factor of the amplitudes (\ref{eq:amp_Ap}) and (\ref{eq:amp_Ap2}), one can rewrite
$
 \tilde{R}~(m_{\Ap})~=~\varepsilon^2~R~(m_{\Ap})
$
and thus finds an upper bound for allowed values of $\varepsilon^2$ as:
\begin{align}
 \varepsilon^2 &< \frac{2\cdot 10^{-6}}{R (m_{\Ap})}.\label{eq:eps_extr}
\end{align}
In Fig.~\ref{fig:decay_rate_plots} (upper and middle panel) the differential decay rate for the signal process relative to the decay $K^+\rightarrow \mu^+ \nu_\mu$ is shown calculated within model I and II for the full phase space (left panels) and with applied corrections due to the given detector acceptance (right panels), according to the experimental set-up of Ref.~\cite{Pang:1989ut}. One notices that within the kinetic mixing model (upper panels of Fig.~\ref{fig:decay_rate_plots}) the internal bremsstrahlung contribution completely dominates the result for the considered $\Ap$ mass parameters: comparison between IB curves and curves including the form factor dependence which was evaluated according to Refs.~\cite{Bijnens:1992en,Poblaguev:2002ug}. Since in model II the gauge invariance is not required, the decay rate is enhanced by a factor of $1/m_{\Ap}^2$ compared to model I. The expected SM background from the decay $K^+\rightarrow\mu^+\nu_\mu \nu\bar{\nu}$ with the applied experimental cuts of Ref.~\cite{Pang:1989ut} is shown in the lower left panel.\\
As one can see from the lower right panel of Fig.~\ref{fig:decay_rate_plots}, the total $\Ap$ decay rate (model II) $\Gamma(K^+\rightarrow\mu\nu_\mu \Ap)$ calculated with $\varepsilon^2=1$ is about a factor of $10^9$ larger than the decay rate to SM particles $\Gamma(K^+\rightarrow\mu^+\nu_\mu \nu_l \bar{\nu}_l)$. This corresponds to an $\Ap$ signal, which will dominate over the expected SM signal for mixing factors down to $\varepsilon^2\simeq10^{-9}$.\\

\begin{figure*}[t!]
 \centering
 \includegraphics[width=.62\linewidth]{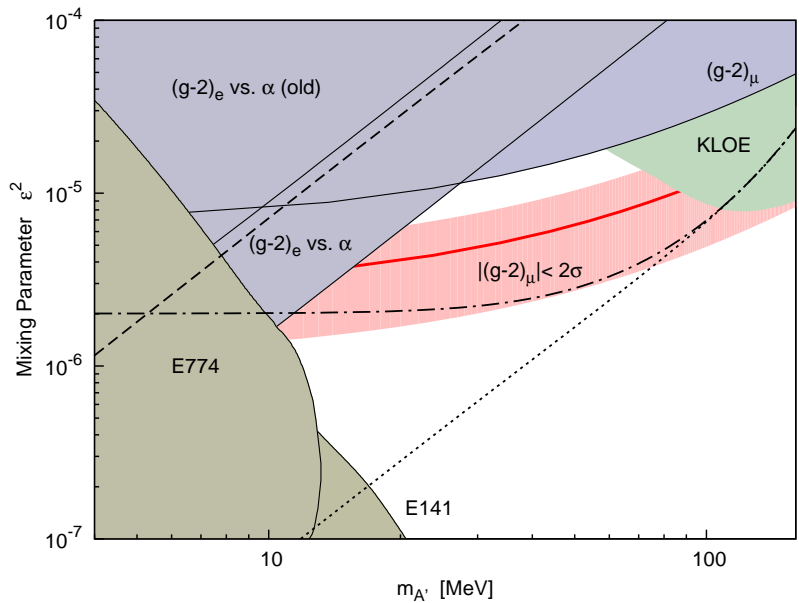}
 \caption{Exclusion limits on the $\Ap$ parameter space. Dashed-dotted curve: bound calculated in the kinetic mixing model (model I) for an accuracy of the ratio $\Gamma(K^+\rightarrow \mu^+ \nu_\mu \Ap)/\Gamma(K^+\rightarrow \mu^+ \nu_\mu)$ of $10^{-9}$. Dashed curve: result for the 1973 data \cite{Pang:1989ut} within model II, where the $\Ap$ only couples to the $\mu^+$. Dotted curve: bound calculated in model II for an assumed improvement of the experimental accuracy by two orders of magnitude, i.e. $2\cdot 10^{-8}$.}
 \label{fig:excl_lim}
\end{figure*}

The calculated limits on the $\Ap$ parameter space are shown in Fig.~\ref{fig:excl_lim}. In this figure the colored regions correspond to already excluded configurations of mass and coupling strength \cite{Bjorken:2009mm,Archilli:2011zc,Blumlein:2011mv}. In this plot we have included the old \cite{Pospelov:2008zw} as well as the new \cite{Aoyama:2012wj,Davoudiasl:2012ig} exclusion limits from $(g-2)$ of the electron compared to the fine structure constant $\alpha$. The red colored region represents the so-called $(g-2)_\mu$ welcome band, where the $\Ap$ contribution to $(g-2)_\mu$ could be invoked to explain the existing discrepancy \cite{Pospelov:2008zw}.\\

A possible bound for the kinetic mixing model is represented by the dash-dotted curve for an assumed experimental accuracy of the ratio $\Gamma(K^+\rightarrow \mu^+ \nu_\mu \Ap)/\Gamma(K^+\rightarrow \mu^+ \nu_\mu)$  of $10^{-9}$. Based on the old $(g-2)_e$ exclusion limit, the 1973 data \cite{Pang:1989ut} allow to slightly improve the bound at low masses and large $\varepsilon$ (dashed curve) within model II. Due to the refinement of the theoretical determination of $(g-2)_e$ the bound from rare kaon data is already covered by the new $(g-2)_e$ limit. Furthermore in Fig.~\ref{fig:excl_lim} we give an estimate in which way the exclusion limits change due to an improvement in the experimental accuracy of the ratio $\Gamma(K^+\rightarrow \mu^+\nu_\mu \Ap)/\Gamma(K^+\rightarrow \mu^+ \nu_\mu)$ by two orders of magnitude (dotted curve). Obviously an improvement of the experimental quantities on the right side of Eq.~(\ref{eq:eps_extr}) will allow the exclusion of a large region of the parameter space up to masses of about 80 MeV since the bound on $\varepsilon^2$ is depending linearly on the RHS of Eq.~(\ref{eq:eps_extr}). Larger angular and momentum acceptancies and a larger rate of stopped $K^+$ compared to \cite{Pang:1989ut} for example will allow to improve this quantity significantly. Such an improved extraction might be achieved by new facilities, such as the NA62 experiment at CERN or rare kaon decay experiments at JPARC.\\

We have used the rare kaon decay $K^+\rightarrow \mu^+ \nu_\mu \Ap$ to find a bound on the $\Ap$ parameters extending the excluded region at low masses in a model where the $\Ap$ couples only to the muon. We have shown, that the method used in this work may be suited to extend the existing limits within two models for the $\Ap$ coupling. For that purpose more precise data are necessary. Improving the accuracy compared to that of the forty year old pioneering work of Ref.~\cite{Pang:1989ut} by two ore more orders of magnitude would allow the exclusion of a significantly larger part of the up to now allowed parameter space, which is also containing a considerable part of the $(g-2)_\mu$ welcome band.\\
This work was supported in part by the Research Centre ``Elementarkraefte und Mathematische Grundlagen" at the Johannes Gutenberg University Mainz and in part by the Deutsche Forschungsgemeinschaft DFG through the Collaborative Research Center ``The Low-Energy Frontier of the Standard Model" (SFB 1044). The authors like to thank Carl Carlson and Achim Denig for helpful discussions, and Bill Marciano for pointing out the gauge invariance constraint within the kinetic mixing model. 

\end{document}